\documentclass[twocolumn,showpacs,superscriptaddress]{revtex4}
\usepackage{amsmath}
\usepackage{latexsym}
\usepackage{amsthm}

\usepackage{amssymb}
\usepackage{graphicx,epstopdf,subfigure,color,hyperref,enumerate}
\usepackage[normalem]{ulem}

\definecolor{Red}{rgb}{1,0,0}

\def\bra#1{\mathinner{\langle{#1}|}}
\def\ket#1{\mathinner{|{#1}\rangle}}
\def\braket#1{\mathinner{\langle{#1}\rangle}}

{\catcode`\|=\active
  \gdef\Braket#1{\begingroup
\mathcode`\|32768\let|\BraVert\left<{#1}\right>\endgroup}}
\def\BraVert{\egroup\,\mid\,\bgroup}

\def\braket#1#2{\mathinner{\langle{#1}|{#2}\rangle}}
\def\es{\mathcal{S}}

\def\ems{{\mathcal{N}}}
\def\emw{{\mathcal{M}}}
\def\eW{\mathcal{W}_g}

\def\as{{\mathcal{A}}}
\def\bs{{\mathcal{B}}}
\def\Ha{{{\mathcal{H}_A}}}
\def\Hb{{{\mathcal{H}_B}}}
\def\Hs{{\mathcal{H}_\es}}
\def\Os{{\Omega}}
\def\sh{{\mathcal{H}}}

\def\W#1{\{#1\}_w}
\newtheorem{prop}{Property}
\newtheorem{proposition}{Proposition}

\begin{document}
\title{Non-local  Measurements via Quantum Erasure}
\author{Aharon Brodutch}
\email{aharon.brodutch@uwaterloo.ca}
\affiliation{Institute for Quantum Computing and Department
of Physics and Astronomy, University of Waterloo, Waterloo, ON,
ON N2L 3G1, Canada}
	
\author{Eliahu Cohen}
\email{eliahuco@post.tau.ac.il}
\affiliation{School of Physics and Astronomy, Tel Aviv University, Tel Aviv 6997801, Israel}
\affiliation{H.H. Wills Physics Laboratory, University of Bristol, Tyndall Avenue, Bristol, BS8 1TL, U.K}

\begin{abstract}

Non-local observables play an important role in quantum theory, from Bell inequalities and various post-selection paradoxes to quantum error correction codes. Instantaneous measurement of these observables is known to be a difficult problem, especially when the measurements are projective. The standard von Neumann Hamiltonian used to model projective measurements cannot be implemented directly in a non-local scenario and can, in some cases, violate causality. We present a scheme for effectively generating the von Neumann Hamiltonian for non-local observables without the need to communicate and adapt. The protocol can be used to perform weak and strong (projective) measurements, as well as measurements at any intermediate strength. It can also be used in practical situations  beyond non-local measurements.  We show how the  protocol can be used to probe a version of Hardy's paradox with both weak and strong measurements. The outcomes of these measurements provide a non-intuitive picture of the pre- and post-selected system. Our results shed new light on the interplay between quantum  measurements, uncertainty, non-locality, causality  and determinism.

\end{abstract}
\maketitle

Many  fundamental questions in quantum mechanics concern measurements and their effects. Much progress has been made regarding the measurability of various, formally defined, `observables' under realistic constraints, with a special emphasis on relativistic and temporal constraints \cite{Beckmanetal2001, AharonovAlbert81, Aharonov1984, AAV1986,Lake2015}, but many questions remain open. In light of relativistic constraints, it is known that measurements cannot violate causality; this limits the types of instantaneous  projective measurements that can be made on spacelike separated systems \cite{PopescuVaidman94, Clark2010}.  Such instantaneous measurements are of interest for a number of reasons.  From a fundamental perspective, we are often interested in space-like separated  subsystems, such as EPR pairs, where communication would rule out the  non-local aspect of an  argument.  From a practical perspective, we want to avoid adaptive schemes, even at the cost of non-deterministic protocols, e.g. linear optics schemes with post-selection  \cite{Kok2007}.

While only few non-local observables can be measured instantaneously with a projective measurement  \cite{AharonovAlbert81}, many others can be measured in a destructive way \cite{GroismanReznik2002, Vaidman2003}. The latter schemes produce the desired probabilities for the outcomes of the measurement. However, they give an unfavorable information gain - disturbance trade-off and usually have a random state at the output, independent  of the input state and measurement result. In this letter we present the \emph{erasure scheme} for effectively creating  the von Neumann measurement Hamiltonian  for a large class of non-local and other non-standard  observables. It  can be used for making strong projective (L\"{u}ders \cite{KirkpatrickLeipzig})   measurements,  weak measurements and measurements at any intermediate strength. Although it  can be used for measuring a wide verity of observables,  we focus on non-local product observable due to their significance.

 We call an operator $\Os$ on a bipartite system (or Hilbert space) $\Hs=\Ha\otimes\Hb$  a \emph{non-local product observable}  when $\Omega=A\otimes B$ and $A,B$ are Hermitian operators on $\Ha$ and  $\Hb$ respectively. We usually consider two observers,  Alice ($\as$) and Bob ($\bs$), with access to $\Ha$ and $\Hb$ respectively, such that in the relevant time interval, $\as$ and $\bs$ are spacelike separated. As a consequence the subsystems cannot interact and $\Omega$ cannot be measured directly. We call the instantaneous  measurement of an observable on a spacelike separated system a \emph{non-local measurement}.

 Non-local product observables play a significant role in quantum theory, for example CHSH observables \cite{PhysRevLett.23.880}, semi-causal measurements \cite{Beckmanetal2001}, non-locality without entanglement \cite{Bennettetal99} and  stabilizer codes \cite{Gottesman2007}.  In some cases, such as the CHSH experiment, it is sufficient to extract the result by making local measurements of $A$ and $B$. Such local measurements disturb the system more than the ideal non-local measurement (see App. \ref{AppA} for examples) and cannot be used in other cases such as quantum error correction and state discrimination, where the outgoing state is as important as the result.  In the case of  weak measurements, the correlations between local measurements are of second order and a  local method does not give the desired result \cite{Brodutch2009,BrodutchMsc}. Weak measurements  of non-local product observables also play an important role in our understanding of quantum mechanics. Examples include Bell tests \cite{Higgins2015,Marcovitch2010, ACE2015}, non-locality via post-selection \cite{AharonovCohen2015} and the quantum pigeonhole principle \cite{AharonovCohen2015,Aharonov2014}. They also play a role in other scenarios such as quantum computing \cite{Lund2011}. Here we demonstrate their significance with  a variant of Hardy's paradox \cite{Hardy1992}. Despite various attempts to find a scheme for non-local  measurements  with a weak limit \cite{Brodutch2009,Resch2004a,Kedem2010} the erasure scheme below is the first scheme that  has both a weak and a strong limit for a wide variety of non-local and other general observables.

\vspace{5pt}
\paragraph*{ \bf The von Neumann scheme:} The standard quantum mechanical  model for a measurement was introduced by  von Neumann and later improved by L\"{u}ders  \cite{KirkpatrickLeipzig} for  degenerate observables. To measure  an observable $\Omega^\es$ on a system  $\es$ we need to couple it to a second quantum system, the \emph{meter} $\emw$, that will register the result of the measurement by the shift of a pointer variable $Q^\emw$ \footnote{Note that the measurement process does not include the readout stage, i.e. it is a coherent process fully described by the quantum dynamics. This is sometimes referred to as a pre-measurement.}. The coupling Hamiltonian is
\begin{equation}\label{vNHam}
H_I=f(t)\Os^\es P^\emw,\end{equation}
where $P^\emw$ is the conjugate momentum to $Q^\emw$ and $f(t)$ is usually an impulse function which is non-vanishing only around the time of the measurement. The interaction strength is $g=\int_0^{\tau} f(t)dt$.  While formally one can write this Hamiltonian for any Hermitian operator $\Omega^\ems$ on $\es$, it may be impossible to implement physically, e.g. when $\Omega^\es$ is a non-local product observable. It is, however, possible to replace the unitary evolution $U=e^{ig\Os^\es P^\emw}$ with an isometry  $V$ such that, for a fixed initial meter state $\ket{0}^\emw$  we get  $V\ket\psi^\es\ket0^\emw=U\ket\psi^\es\ket0^\emw=\sum_ka_k\ket{k}^\es\ket{\lambda_k}^\emw$, where  $\ket{\psi}^\es=\sum_ka_k\ket{k}^\es$  is an arbitrary system state, and  $\ket{k}^\es$ are eigenstates of $\Os^\es$, $\Os^\es\ket{k}^\es=\lambda_k\ket{k}^\es$.  While the implementation of  $V$ induces the desired dynamics, it may have two drawbacks: First it may depend on the initial state of the meter, second it might not have a free parameter corresponding to the measurement strength $g$.  Both  appear in standard non-local measurement schemes such as modular measurements \cite{AharonovAlbert81}.

After the measurement, the   system  state is dephased in the eigenbasis of $\Omega^\es$, however if $\Omega^\es$ is degenerate, each degenerate subspace remains coherent. The measurement  is usually followed by reading out the state of the pointer $Q^\emw$. When the shift in $Q^\emw$ is large compared to the uncertainty $\Delta_Q$, i.e. $\braket{\lambda_k}{\lambda_l}\approx \delta(\lambda_k-\lambda_l)$, the measurement is \emph{strong} and the result of a single measurement is unambiguous, thus dephasing is complete. When the  possible shift in $Q$ is much smaller than $\Delta_Q$,  we have a weak measurement.  Within the von Neumann model this can be achieved by choosing the coupling strength  $g$ to be small enough, or by increasing $\Delta_Q$. As a result of the weak measurement,  the system is only slightly dephased.

Weak measurements allow us to ask questions about a system at an intermediate time between an initial preparation of the state  $\ket{\psi}$ (pre-selection), and a final projective measurement leading to $\ket{\phi}$ (post-selection),  without making counterfactual statements.
The result is a complex number called the \emph{weak value},
\begin{equation}\label{eq:wv}
\{\Os\}_w=\frac{\bra\phi \Os \ket\psi}{\braket\phi\psi}.
\end{equation}
Although the readout requires many identical experiments, in each experiment the result is encoded in a quantum  meter whose dynamical evolution is dictated by a weak potential term in the  Hamiltonian  $H_w=\{\Os\}_wP^\emw$  \cite{Aharonov2014Unusual}.

\paragraph*{ \bf Quantum erasure:}   A  description of a   \emph{quantum eraser} \cite{Scully82,Herzog95,Walborn02,Peruzzo12} is simple when the meter has a discrete  Hilbert space and  $\{\ket{\lambda_k}^\ems\}$ is an orthonormal basis.  Before the readout stage it  is  possible to undo or \emph{erase}  the measurement locally in $\ems$ by measuring in the conjugate   basis to $\{\ket{\lambda_k}^\ems\}$    and post-selecting the result corresponding to the POVM element $\Pi_{+M}^\ems=\ket{+M}\bra{+M}^\ems$, with $\ket{+M}^\ems=\sum_k\ket{\lambda_k}^\ems$,
\begin{equation}
\Pi_{+M}^\ems\left[U\ket{\psi}^\es \ket{0}^\ems\right] \propto \ket{\psi}^\es\ket{+M}^\ems.
\end{equation}
The erasure procedure is probabilistic, but  we can make it deterministic by considering all POVM elements and adding a unitary operation at the end.

Note: A scheme for erasing weak measurements  \cite{Scully2011} and a relation between weak measurement and erasure \cite{Tamate2009} were recently proposed. In contrast, our method below utilizes the quantum erasure of a strong measurements as a tool for performing general measurements.

\vspace{5pt}
\paragraph*{ \bf Main result:} The \emph{erasure scheme} below involves  two meters: $\ems$ and $\emw$.  The pointer  for $\ems$ is  $Q^\ems$ so $Q^\ems\ket{\lambda_k}^\ems=\lambda_k\ket{\lambda_k}^\ems$, likewise $Q^\emw$ is the pointer for $\emw$.

{\proposition{It is possible to induce the von Neumann coupling between  $\es$ and $\emw$  by making  a strong  measurement of $\es$ with $\ems$ and erasing the result.}\label{prop1}}

\proof Let $\sum_ka_k\ket{k}^\es\ket{\lambda_k}^\ems$ be the system meter state after a strong measurement.  The second meter $\emw$ is in the arbitrary initial state $\ket{0}^\emw$. We now let $\emw$ interact with $\ems$ using the unitary $e^{igQ^\ems P^\emw}$
\[e^{igQ^\ems P^\emw}\sum_ka_k\ket{k}\ket{\lambda_k}\ket{0}=\sum_ka_k\ket{k}\ket{\lambda_k}e^{ig\lambda_k P^\emw}\ket{0},\]and then erase using $\Pi_{+M}$
\[\Pi_{+M}\sum_ka_k\ket{k}\ket{\lambda_k}e^{i\lambda_kP^\emw}\ket{0}\propto\sum_ka_k\ket{k}\ket{+M}e^{ig\lambda_kP^\emw}\ket{0}.\]
This is the dynamics induced by the Hamiltonian \eqref{vNHam}. \qed

 In the following we show how this method can be used for measurements of non-local product observables.

\begin{figure}[h]
\includegraphics[width=0.98\columnwidth]{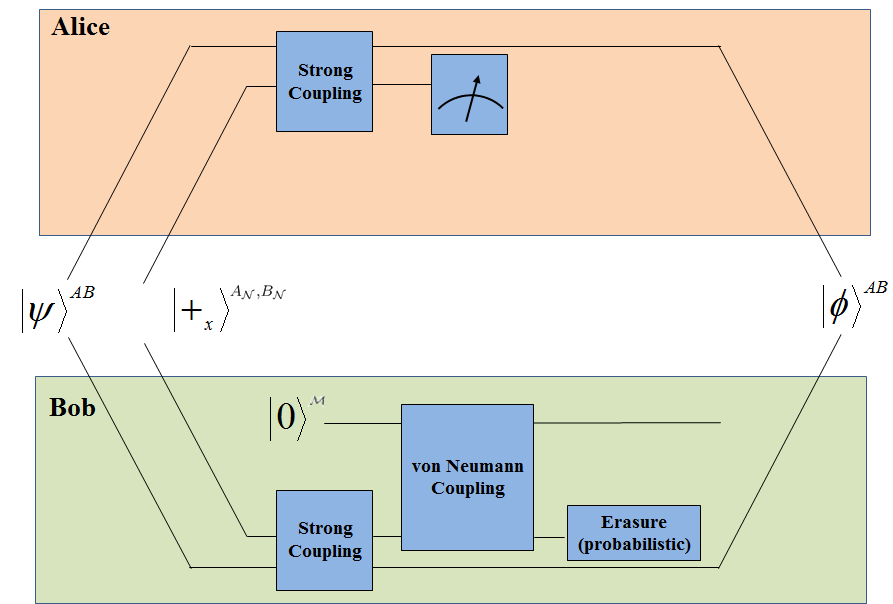}
\caption{{\bf Non-local  measurement based on quantum erasure.} The strong measurement   requires Alice and Bob to locally couple their system to an entangled meter  (the ancilla $\ems$). Alice then  measures her part, $\ems_A$,  effectively  `pushing' the result of the strong measurement to  Bob's $\ems_B$.  Bob performs the weak measurement of  $\ems_B$ using $\emw$ and then erases the information encoded on $\ems$ (undoing the initial coupling).  If successful they induce the von Neumann Hamiltonian \eqref{vNHam} \label{fignl}}
\end{figure}

\vspace{5pt}
\paragraph*{ \bf  Measurement of  product observables:} The challenge  with measuring a non-local observable is to couple to the degenerate subspaces according to  the  L\"{u}ders rule.  Given a bipartite system 
and two local Hermitian operators, $X$ on $\Ha$ and $Y$ on $\Hb$, the degenerate subspaces of  $XY$  are generally different from those of  the local observables $X$ and $Y$.  The erasure procedure above can be used to remove the redundant information encoded  locally.  Below we combine this method with a remote measurement to produce  the Hamiltonian \eqref{vNHam} with   $\Os^\es\rightarrow XY$ (see Fig. \ref{fignl}).

$\ems$ is prepared in an entangled state on the Hilbert space $\sh_\ems=\sh_{A_\ems}\otimes\sh_{B_\ems}$, that  depends on the properties of $X$ and $Y$.  $\emw$ is local at Bob's side and has an initial state $\ket{q=0}$.  Alice locally couples the entangled $\ems$ to her subsystem, performing a strong measurement with the result encoded non-locally (i.e. Alice cannot access the result alone). Alice then reads out the state of her strong meter. This teleports the result to Bob (possibly with a known offset). Next, Bob performs the procedure outlined  in the proof of proposition \ref{prop1}. The resulting dynamics is $U=e^{igXYP}$.

\vspace{5pt}

\emph{Details}: We define the sets of orthogonal projectors $\{\hat X_k\}$, $\{\hat Y_l\}$ such that $X=\sum_{k=1}^{|x|}x_k\hat X_k$ and $Y= \sum_{l=1}^{|y|}y_l\hat Y_l$, where $\{x_k\}$ and $\{y_l\}$ are the sets of  distinct  eigenvalues of $X$ and $Y$ with cardinality $|x|$ and $|y|$ respectively.  Assume (without loss of generality) that $|x|\le|y|$.  $\ems$ will have dimension $|x|\times|x|$ and will be prepared in the initial entangled state $\ket{+_x}^\ems=\frac{1}{\sqrt{|x|}}\sum_{m=0}^{|x|-1}\ket{m}^{A_\ems}\ket{m}^{B_\ems}$. The system is initially in the  unknown state $\ket{\psi}=\sum_{i,k}\alpha_{i,k}\ket{i,k}^{A,B}$.  We denote the global ($A,B,\ems,\emw$) initial state by $\ket{\Psi_0}$. We also define $K_\mu=\sum_k (k-\mu) X_{k}$. The scheme is as follows:

1. Alice  couples  $K_0$ and  $\ems$ using $U_s\ket{i}\ket{m}=\ket{i}\ket{m+i}$, producing  \[
\ket{\Psi_1}=\sum_m\sum_{i,k}\alpha_{i,k}\ket{i,k}^{A,B}\ket{m+i,m}^{A_\ems,B_\ems}\ket{q=0}.\]

2. Alice  reads out  $\ems^A$ and gets a result corresponding to $\ket{\mu}$. Thus
\begin{equation}
\ket{\Psi_2}=\sum_{i,k}\alpha_{i,k}\ket{i,k}^{A,B}\ket{\mu}^{A_\ems}\ket{\mu-i}^{B_\ems}\ket{q=0}\end{equation}
Note that the label $\mu-i$ is modular, i.e. $\ket{\mu-i}^{B_\ems}=\ket{\mu-i\pm |x|}^{B_\ems}$.

 3. Bob  now has  access to the operator $K_\mu$, so  he can  couple to $\Omega_\mu^\ems=[\sum_k x_{k-\mu} X_{k}]\otimes Y$  using the local interaction  Hamiltonian  $Y^BQ^{B_\ems}P^\emw$, where $Q^{B_\ems}\ket{\mu-i}^{B_\ems}=x_{i-\mu}\ket{\mu-i}^{B_\ems}$.

4. Bob  erases  Alice's measurement with probability $1/|x|$. If he succeeds, the effective $\es-\emw$  dynamics is $U=e^{ig\Omega_\mu^\es P^\emw}$.

For $\mu=0$ this is  the desired observable, and in some special cases it  is a simple  re-scaling for all $\mu$ (see App. \ref{AppC}). The worst case measurement  will succeed with probability $1/|x|^2$ (both erasure  and $\mu=0$ are required) while the best case will succeed with probability $1/|x|$ (only erasure  is required). In either case failure would correspond to a non-trivial (but known) unitary evolution during the interval between pre- and post-selection.
For a more detailed  description see App. \ref{AppC}.

\vspace{5pt}
\paragraph*{ \bf Determinism and non-locality:} The protocol is probabilistic, however it can be turned into a deterministic protocol if Alice and Bob are allowed to communicate. This is to be expected since the von Neumann Hamiltonian of a product operator measurement (or even  the less general isometry $V$)  can be used for signalling  between Alice and Bob \cite{Clark2010,PopescuVaidman94}. The entanglement and communication resources for our scheme are at most equivalent to  a single round of teleportation. This can be compared to the naive strategy of  teleporting, measuring and teleporting back. In the example below,  and the one in App.  \ref{example2} ,  we show  that the communication cost of our scheme saturates the lower bound imposed by causality.

However, the motivation for the protocol is the fact that it can be implemented without communication or  adaptive components.  The non-local paradox below is a good example of a situation where communication is not allowed by assumption,  as is the case with the  Bell inequality. From a practical perspective we can easily imagine other  situations such as  linear optics, where the resources  necessary for an adaptive scheme that requires communication outweigh the advantage of a deterministic protocol \cite{Kok2007}.

In a post-selected scenario  with a future boundary condition $\ket\phi^{A,B}$, it is possible to include the post-selection requirement for the measurement in the future boundary conditions.  The pre-selected system would then be $\ket{\psi}^{A,B}\ket{+_x}^{A_\ems,B_\ems}$, the post-selection would be $\ket{\phi}^{A,B}\ket{0}^{A_\ems}\ket{+M}^{B_\ems}$.
Taking  $U_s$ into account gives,  \[\{YQ^{B_\ems}\}_w=\frac{\bra{+M}\bra{0}\bra{\phi}U_s^\dagger YQ^{B_\ems}\ket{\psi}\ket{+_x}}{\bra{+M}\bra{0}\bra{\phi}U_s^\dagger \ket{\psi}\ket{+_x}}=\{XY\}_w.\]

\vspace{5pt}
\paragraph*{\bf Example:}
Consider a two qubit system, where  Alice and Bob each have local access to a single qubit. The observables of interest are  the  local projectors  $\Pi_{m,\cdot}=\ket{m}\bra{m}\otimes\openone$,  $\Pi_{\cdot,m}=\openone\otimes\ket{m}\bra{m}$  and  the non-local projector $\Pi_{m,n}=\Pi_{\cdot,m}\Pi_{n,\cdot}=\ket{m}\bra{m}\otimes\ket{n}\bra{n}$ with $m,n\in\{0,1\}$.  Now let $\emw$ be a meter with conjugate momentum $P^\emw$ located on Bob's side. Our scheme allows us to create the effective interaction Hamiltonian $H=\Pi_{m,n}P^\emw$  with probability $\frac{1}{4}$, the maximal probability allowed by causality constraints (see App. \ref{AppC}).

\begin{figure}[h]
\includegraphics[width=0.98\columnwidth]{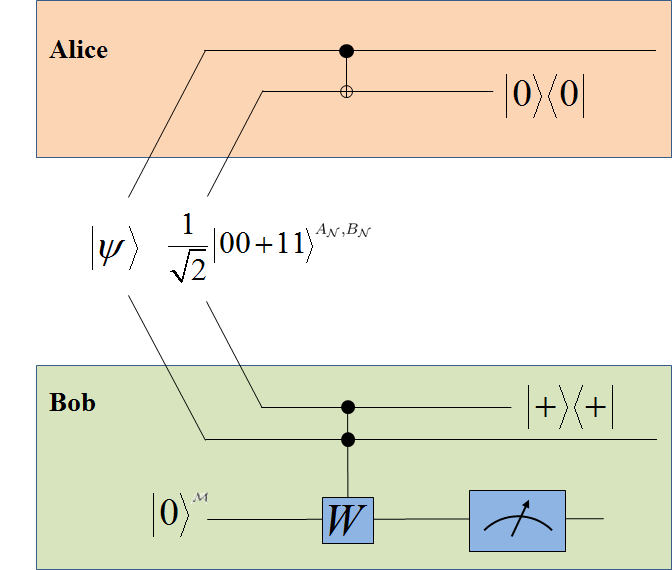}
\caption{{\bf Instantaneous measurement of $\Pi_{1,1}$}  \label{figpi}. To measure the non-local observable $\Pi_{1,1}=\ket{11}\bra{11}$ Alice and Bob need an entangled ancilla  $\ems$ and a local meter  $\emw$ (on Bob's side).  They locally couple using a CNOT on Alice's side and a controlled-controlled-$W$ on Bob's side with $W=e^{igP^\emw}$. The value of $g$ determines the measurement strength. After a successful (local) post-selection of the state $\ket{0+}$ for the ancilla the effective system-meter coupling will be $e^{ig\Pi_{1,1}P^\emw}$. }
\end{figure}

The explicit scheme is as follows (see Fig. \ref{figpi}):  The ancilla $\ems$ is  prepared in the entangled state $\ket{+_x}=\frac{1}{\sqrt{2}}\ket{00+11}^{A_\ems,B_\ems}$, $U_s$ is  a CNOT between $A_\ems$ and Alice's subsystem. The interaction with $\emw$ is  a Controlled-Controlled-$W$  between $B,B_\ems$ and $\emw$ , with $W=e^{igP^\emw}$. Following the interactions,  Alice and Bob post-select the state $\ket{0}^{A_\ems}\ket{+}^{B_\ems}$ on the ancilla (with probability $\frac{1}{4}$).  The induced transformation is $U=e^{ig\Pi_{1,1}P^\emw}$.   Bob can choose $g$ to make the measurement weak or strong.

In principle, Alice and Bob do not need to coordinate their actions. They can each freely choose which operator to couple  without notifying the other. Moreover, Bob can choose  $W=e^{igP^\emw}$ without notifying Alice.

\vspace{5pt}
\paragraph*{\bf A non-local paradox:}

In the  EPR scenario, a bipartite  system has a definite state with respect to a non-local observable but has random marginals \cite{Schrodinger1935}. In a post-selected regime it is possible to observe the opposite behavior, i.e. a system with definite local properties but uncertain non-local ones. Let  $\ket{\psi_H}=\frac{1}{\sqrt{2}}[\ket{0}\ket{-}-\ket{1}\ket{+}]$  be a pre-selected state and $\ket{\phi_H}=\ket{+}\ket{+}$  be the post-selection. 
 If either Alice or Bob make a local measurement of $\Pi_{1,\cdot}$ or $\Pi_{\cdot,1}$ respectively, they will expect the result $1$ with certainty. This follows from the Aharonov, Bergman and Lebowitz (ABL) formula for calculating probabilities on pre- and post-selected systems \cite{Aharonov1964}. If this were a classical scenario, it  would have implied that  a measurement of $\Pi_{1,1}$ should also produce the outcome $1$ deterministically. However, the probability of obtaining  the outcome $1$ for a measurement of $\Pi_{1,1}$ is $\frac{1}{2}$.

The scheme presented in the previous section allows us to directly measure $\Pi_{1,1}$.  If instead  we measure $\Pi_{1,1}$ indirectly via $\Pi_{1,\cdot}$ and $\Pi_{\cdot,1}$, we will get the results $1,1$ with probability $\frac{1}{4}$ (see App \ref{AppB} for details).

One  may see  the paradox as a result of measurement disturbance.  Weak measurements let  us  avoid this issue. The  local  weak values are   $\{\Pi_{1,\cdot}\}_w=\{\Pi_{\cdot,1}\}_w=1$, while the non-local one is   $\{\Pi_{1,1}\}_w=\frac{1}{2}$. Here  we see the full power of our scheme. It  allows the first direct  measurement of these weak values.

Non-local weak   measurements  were previously used to provide an elegant solution to Hardy's paradox \cite{Hardy1992,Aharonov2002}. The same logic applies in the example of above. The non-local weak values are  $\{\Pi_{1,1}\}_w=\{\Pi_{0,1}\}_w=\{\Pi_{1,0}\}_w=-\{\Pi_{0,0}\}_w=\frac{1}{2}$. The last weak value is negative and ensures the weak values add up to  $1$.  In Hardy's experiment it can be associated with negative occupation numbers in an interferometer.  Using Pusey's construction \cite{Pusey2014} it is possible to show that the negative weak value is a result of  contextuality. In this case the context is the information of the measurement regarding local observables.



\vspace{5pt}
\emph{\bf Generalizations\label{comp}}: It is possible  to generalize the erasure scheme to other types of operators $\Omega$.  If the measured operator is separable, like the Bell operator, it is possible to measure each product operator and add the results on a single meter \cite{Brodutch2009}. It is also possible to perform more general measurements of a degnerate observable by decomposing the  measurement into extremal POVMs \cite{Sentis2013}and using  erasure technique  to  coarse-grain the outcome.  Another class of measurable operators  are non-Hermitian operators  resulting from  sequential  measurements.  These  measurements are  natural in various settings such as tests of contextuality and Leggett-Garg inequalities and measurements  of  quantum trajectories \cite{Marcovitch2010,FHofmann2014,Mermin1993,Mitchison2007}. The specifics of an erasure based sequential measurement scheme are given  elsewhere \cite{BCpath}.

Finally, measurements are only one possible application of the Hamiltonian \eqref{vNHam}. The  erasure scheme can be modified to generate this   Hamiltonian   under a wide set of constraints that prevent the direct coupling of a system $\emw$ to a degenerate operator  $\Omega^\es$.   In App. \ref{AppE}, we show how to use the erasure technique to construct a generic Controlled-Controlled-Unitary gate in cases where the relevant qubits cannot interact, a common restriction for photonic qubits.

\vspace{5pt}
\emph{\bf Conclusions}:     We presented the \emph{erasure scheme} for effectively creating a von Neumann measurement Hamiltonian for a non-local observable. It  is based on the fact that it is possible to perform a von Neumann measurement by making a strong measurement and erasing the result (Proposition \ref{prop1}).

The scheme  has a number of advantages over known schemes for instantaneous non-local measurements: It can be used in the strong, intermediate and weak regimes; so far this range was only possible for the special case of sum observables \cite{Brodutch2009}.  It is also versatile in terms of the types of observables that can be measured, other schemes such as the modular measurement \cite{AharonovAlbert81}, Kedem-Vaidman \cite{Kedem2010} and Resch-Steinberg (RS) \cite{Resch2004a} schemes can only be used for  specific subsets of non-local observables and cannot be further generalized (see App. \ref{AppD} for details).  Another advantage is that it can be used for a much wider class of observables than those presented here, for example multipartite observables and non-Hermitian operators \cite{BCpath}.

Causality constraints imply a probabilistic scheme. However, with clever post-processing and correction techniques it can be used to get a result on every  possible run. In some cases, causality constraints rule out the possibility that the outcome would always correspond to the desired observable.  `Failing' post-selection would either give a result for  a different operator and/or act like a known unitary in the intermediate time between the pre- and post-selection.   It is possible to avoid these constraints in post-selected scenarios  by including the probabilistic element in the post-selection.
The limitations of the scheme, and the possible ways to overcome them, demonstrate the subtle interplay between causality, determinism and quantum measurement.

The scheme has many potential uses. Here we highlighted its role  in tests of quantum foundations in non-local scenarios.  In a future publication \cite{BCpath} we will show that the scheme can be used for sequential experiments such as those used in tests of contextuality and Leggett-Garg inequalities. In these sequential scenarios the measurement is not instantaneous but the causality constraints are stricter since they explicitly involve communication backwards in time.

Regarding experimental realizations, the scheme is feasible in optics and other platforms such as NMR \cite{Lu2013} and atomic spontaneous emission \cite{Shomroni2013}. For a weak measurement it  has a significant advantage over the RS scheme \cite{Resch2004a, Lundeen2009} since the resources required are linear in $g$ as opposed to the RS scheme that scales quadratically (see App. \ref{AppD} for details).   It would be interesting to see if current methods can be used to perform the full version of these techniques including the correction and post-processing steps. It  would also be interesting to find  further applications for the erasure method such as   improved experimental accuracies \cite{Dressel2014} or  protective tomography  \cite{money} at the weak limit, error correction at the strong limit \cite{Ghosh2012} or for generating  many-body interactions under realistic constraints (for an example see  App. \ref{AppE}).

\section*{Acknowledgments}
We thank Yakir Aharonov, Raymond Laflamme, Marco Piani, Or Sattath and referee B  for their valuable input on this work. AB is partly supported by NSERC, Industry Canada and CIFAR. EC was supported in part by the Israel Science Foundation Grant No. 1311/14 and by ERC AdG NLST.






\onecolumngrid

\appendix

\section{Local and non local projective measurements of product observables on two qubits}\label{AppA}
Quantum measurements induce a back-action on the measured system. The back-action depends on the observable being measured, as well as the method used to perform the measurement. In the case of a non-local product observables $A\otimes B$, it is possible to design a measurement that  generates the desired statistics for the outcomes by performing local measurements of $A$ and $B$ and post-processing the joint results.   Such a measurement, however, often  induces a back-action which is different (and more destructive) than the corresponding non-local L\"uders measurement of $A\otimes B$.  To illustrate this fact we use two simple examples. In the first example the channel induced by the non-local measurement can be used to entangle the system, consequently the measurement cannot be performed by local operations and classical communication (LOCC) without shared entanglement resources.  In the second example the non-local measurement induces a channel that can  be used to both entangle the system and transmit information. Consequently,  it requires bidirectional communication (i.e. it is not  localizable in the sense of \cite{Beckmanetal2001}) and shared entanglement resources. 

\subsection{Notation}
We consider the measurement of the  operator $\Omega=\sum_k\omega_k\Pi_k$  where $\{\Pi_k\}$ are orthogonal  projection operators and $\{\omega_k\}$ are distinct eigenvalues of $\Omega$. Note that if  $\Omega$ is degenerate, there is at least one $\Pi_k$ of rank $>1$. 

The projective L\"uders measurement of $\Omega$ induces the channel $M$ on the state of the system $\rho$.  
\[M(\rho)=\sum_k \Pi_k \rho \Pi_k.\]
 The measurement on $\rho$ will produce the result corresponding to $\omega_k$ with probability given by the Born rule $P(\omega_k)=tr[\Pi_k\rho]$.  If the measurement produces the result $\omega_k$, the resulting sub-channel is $M_k(\rho)=\Pi_k\rho\Pi_k$. Note that we use the convention where the sub-channel is not trace preserving. 

In what follows we will sometimes use the fact that the sub-channels are projections so that we can use pure state notation. In those cases we  replace the notation $M_k(\ket{\psi}\bra{\psi})$ with 
\begin{equation}
M_k(\ket{\psi})=\Pi_k\ket\psi
\end{equation}

\subsection{Measurements of $\Omega=\sigma_z^A\otimes\sigma_z^B$}

Consider the measurement of $\Omega=\sigma_z^A\otimes\sigma_z^B$. The Krauss operators for the L\"uders channel are $\Pi_{+1}=\ket{00}\bra{00}+\ket{11}\bra{11}$ and $\Pi_{-1}=\ket{10}\bra{10}+\ket{01}\bra{01}$. Now take  an initial separable state $\ket{\psi}=\frac{1}{2}[\ket{0}+\ket{1}]\otimes[\ket{0}+\ket{1}]=\frac{1}{2}[\ket{00}+\ket{01}+\ket{10}+\ket{11}]$. Let us assume we perform the non-local measurement of $\Omega$ and get the result $+1$, the outgoing state will be (up to normalization)
\begin{equation}
\Pi_{+1}\ket\psi=\frac{1}{2}[\ket{00}+\ket{11}]
\end{equation}
which is  entangled. 

\subsubsection{The corresponding local measurement}
Consider now the joint measurement broken into  two local measurements $\sigma_z^A$  and  $\sigma_z^B$. The joint measurement has four possible outcomes $\{(+1,+1),(+1,-1),(-1,+1),(-1,-1)\}$, it induces a channel $\tilde M$ with Krauss operators $\tilde\Pi_{+,+}=\ket{00}\bra{00}$, $\tilde\Pi_{+,-}=\ket{01}\bra{01}$, $\tilde\Pi_{-,+}=\ket{10}\bra{10}$ and $\tilde\Pi_{-,-}=\ket{11}\bra{11}$.  Consequently, the sub-channels are entanglement breaking. 

We can coarse grain the measurement by forgetting the local results so that the sub channels will be 
\[\tilde M_{+1}(\rho)=\ket{00}\bra{00}\rho\ket{00}\bra{00}+\ket{11}\bra{11}\rho\ket{11}\bra{11}\]
and 

 \[\tilde M_{-1}(\rho)=\ket{01}\bra{01}\rho\ket{01}\bra{01}+\ket{10}\bra{10}\rho\ket{10}\bra{10}\]
 These sub-channels are  still entanglement breaking. 

\subsection{The  L\"uders measurement $\Omega=\ket{11}\bra{11}$ is not localizable}

A bipartite  channel is said to be localizable (in the sense of \cite{Beckmanetal2001}) if it cannot be implemented using local operations and shared entanglement (in other words,  it cannot be implemented instantaneously). A bipartite channel is called causal if it does not allow transmission of information between the two parts. Causality is a pre-requisite for localizability. In what follows we show that the  L\"uders measurement of $\Omega=\ket{11}\bra{11}$ can be used to transmit information between Alice and Bob, consequently it is neither causal nor localizable. 

Consider the measurement of   $\Omega=\ket{11}\bra{11}$ . The Krauss operators for the channel $M(\rho)$ are $\Pi_1=\ket{11}\bra{11}$ and $\Pi_0=\ket{00}\bra{00}+\ket{10}\bra{10}+\ket{01}\bra{01}$. To show that this channel can produce entanglement, consider the same initial state $\ket{\psi}$ as the previous section. The sub-channel corresponding to the result $0$ will produce an entangled state.

To show that the channel $M(\rho)$ can be used to transmit information, consider the following strategy for Alice to signal Bob. Alice and Bob prepare the initial state $\ket{Zero}=\frac{1}{\sqrt{2}}\ket{0}\otimes[\ket{0}+\ket{1}]$. Now Alice can signal Bob by either doing nothing, in which case $M(\ket{Zero})=\ket{Zero}$, or by changing her state locally to $\ket{1}$ so that global state reads $\frac{1}{\sqrt{2}}\ket{1}\otimes[\ket{0}+\ket{1}]=\ket{One}$. Now 
\begin{equation}
M(\ket{One}\bra{One})=\frac{1}{2}[\ket{10}\bra{10}+\ket{11}\bra{11}]
\end{equation}
 Bob's local state is  $\frac{1}{2}[\ket{0}\bra{0}+\ket{1}\bra{1}]$. What we see is 
\begin{equation}
Tr_A[M(\ket{Zero}\bra{Zero})]\ne Tr_A[M(\ket{One}\bra{One})]
\end{equation}
 Alice can thus use the channel $M$ together with local operations on her side to change Bob's state, consequently the channel is not causal (and not  localizable). 

\section{A non-local paradox}\label{AppB}
Hardy's paradox \cite{Hardy1992} involves an electron and a positron going through two overlapping interferometers.  By clever pre- and post-selection it is possible to observe a seemingly paradoxical situation where the particles appear to  follow  an  impossible trajectory.  In what follows we present a variant of this paradox with two modifications. First, we modify the setting to a more abstract `qubit' setting. Second, we modify the pre- and post-selection to allow a situations where  local measurements have  deterministic outcomes, while the analogous non-local measurements are completely uncertain.  As in Hardy's paradox \cite{Aharonov2002}, the counterfactual reasoning of strong measurements is supported by  weak values.

 Alice and Bob initially share two qubits prepared in the entangled (pre-selected) state $\ket{\psi_H}=\frac{1}{\sqrt{2}}[\ket{0}\ket{-}-\ket{1}\ket{+}]$  and later locally post-selected in the state  $\ket{\phi_H}=\ket{+}\ket{+}$.  The observables to be measured are  $\Pi_{m,\cdot}=\ket{m}\bra{m}\otimes\openone$ ,  $\Pi_{\cdot,m}=\openone\otimes\ket{m}\bra{m}$  and $\Pi_{m,n}=\Pi_{\cdot,m}\Pi_{n,\cdot}=\ket{m}\bra{m}\otimes\ket{n}\bra{n}$ with $m,n\in\{0,1\}$.

The weak values can be  calculated using
\begin{equation}\label{eq:wv}
\{\Os\}_w=\frac{\bra\phi \Os \ket\psi}{\braket\phi\psi}.
\end{equation}
Hence locally
\begin{equation}
\{\Pi_{1,\cdot}\}_w = \frac{\langle\phi_H| 1 \rangle \langle 1 |\otimes\openone|\psi_H\rangle}{\langle\phi_H|\psi_H\rangle}=1,
\end{equation}
and similarly
\begin{equation}
\{\Pi_{\cdot, 1}\}_w = \frac{\langle\phi_H| \openone\otimes | 1 \rangle \langle 1 |\psi_H\rangle}{\langle\phi_H|\psi_H\rangle}=1.
\end{equation}
However, the non-local weak value of $\Pi_{1, 1}$ is
\begin{equation}
\{\Pi_{1, 1}\}_w = \frac{\langle\phi_H| 1 \rangle \langle 1 | 1 \rangle \langle 1 |\psi_H\rangle}{\langle\phi_H|\psi_H\rangle}=1/2,
\end{equation}
and also $\{\Pi_{0,1}\}_w=\{\Pi_{1,0}\}_w=-\{\Pi_{0,0}\}_w=\frac{1}{2}$.

The strong values can be derived using the ABL rule \cite{Aharonov1964}, which states
\begin{align} \label{ABL}
Pr\left(A_{k}|\psi,\phi\right)&=
\frac{\left|\left\langle \phi|A_k|\psi\right\rangle \right|^{2}}{\sum_{j}\left|\left\langle \phi|A_j|\psi\right\rangle \right|^{2}},
\end{align}
where $A$ is a measurement operator characterized by the projectors $A_k$, such that $A=\sum_ka_kA_k$.
The probability for the result $1,1$ is therefore $\frac{1}{2}$ for the non-local measurement, but $\frac{1}{4}$ for the joint local measurements.

\section{Non-local measurement of a general product} \label{AppC}

The aim of the protocol is coupling a non-local system $\es=\as\otimes\bs$ with Hilbert space $\sh_\es=\sh_A\otimes\sh_B$ to a meter $\emw$ with Hilbert space $\sh_\emw$ via the induced unitary $U=e^{iXYP^\emw}$, where $X$ is an operator on $\sh_A$, $Y$ is an operator on $\sh_B$ and $P^\emw$ is an operator on  $\sh_\emw$ .

\subsection{Definitions}
 $\{\hat X_i\}$, $\{\hat Y_i\}$ are sets of orthogonal operators such that $X=\sum_{i=1}^{|x|}x_i\hat X_i$ and $Y= \sum_{k=1}^{|y|}y_k\hat Y_k$ where $x_i$ and $y_i$ are the distinct  eigenvalues of $X$ and $Y$ so  $x_i\ne x_j~\forall i\ne j$, $y_k\ne y_l~\forall k\ne l$. The sets $\{x_k\}$ and $\{y_l\}$ have cardinality $|x|$ and $|y|$ respectively.   We label $\as$ and $\bs$  such that   $|x|\le|y|$.

We use an ancillary  meter $\ems$ with  Hilbert space $\sh_{A_\ems}\otimes\sh_{B_\ems}$ of  dimension $|x|\times|x|$ and initialize it in an entangled state $\ket{+_x}=\frac{1}{\sqrt{|x|}}\sum_{m=0}^{|x|-1}\ket{m}^{A_\ems}\ket{m}^{B_\ems}$. The system $A,B$ is initially in the unknown state $\ket{\psi}=\sum_{i,k}\alpha_{i,k}\ket{i,k}^{A,B}$ and $\emw$ is in the arbitrary initial state $\ket{q=0}$.   We denote the total  initial state by

\[\ket{\Psi_0}=\ket\psi^{A,B}\ket{+_x}^{A_\ems,B_\ems}\ket{q=0}^\emw.\]

$U_s$ is the strong measurement interaction on $A$,  such that
\[U_s\ket{i}^A\ket{m}^{A_\ems}=\ket{i}^A\ket{m+i}^{A_\ems},\]
where the  labels $m$ are modular, i.e. $\ket{m+i}=\ket{m+i\pm|x|}$.

The  states $\ket{m}^{B_\ems}$ are eigenstates of the operator  $Q^{B_\ems}$ with eigenvalues $x_{-m}$, so that  $Q^{B_\ems}\ket{m}^{B_\ems}=x_{-m}\ket{m}^{B_\ems}$. The meter $\emw$ has a position variable $Q^\emw$ and conjugate momentum $P^\emw$.

 We define the operators \[\Omega_\mu^\ems= [\sum_kx_{k-\mu}X_{k}]\otimes Y, \] on $\sh_A\otimes\sh_B$.

Finally,  we use the convention
\[e^{-i{g} P^\emw}\ket{q=0}=\ket{q=g}.\]

\subsection{Protocol}
Initially, the total state is
\begin{align}
\ket{\Psi_0}=\sum_m\sum_{i,k}\alpha_{i,k}\ket{i,k}^{A,B}\ket{m,m}^{A_\ems,B_\ems}\ket{q=0}^\emw.
\end{align}

\subsubsection{Alice's operations}

1. Applying $U_s$ and 2. measuring $A_\ems$ (with result $\mu$) yields

\begin{align}
\ket{\Psi_0}&\rightarrow\ket{\Psi_1}=\sum_m\sum_{i,k}\alpha_{i,k}\ket{i,k}^{A,B}\ket{m+i,m}^{A_\ems,B_\ems}\ket{q=0}^\emw\\
&\rightarrow\ket{\Psi_2}=\sum_{i,k}\alpha_{i,k}\ket{i,k}^{A,B}\ket{\mu-i}^{B_\ems}\ket{q=0}^\emw.\label{eq:p2}
\end{align}

\subsubsection{Bob's operations}

3. Bob uses the coupling Hamiltonian
\begin{align}
H_I=f(t)Y^B Q^{B_\ems}P^\emw.
\end{align}
 With $\int f(t)dt =g$ we get the unitary evolution $e^{-i{g}Y^B Q^{B_\ems}P^\emw}$, which produce the state
\begin{align}
\ket{\Psi_{3}}=\sum_{i,k}\alpha_{i,k}\ket{i,k}^{A,B}\ket{\mu-i}^{B_\ems}\ket{q=g x_{i-\mu}y_k}.
\end{align}

4. Bob erases the result on  $B_\ems$ (effectively undoing Alice's coupling $U_s$) by post-selecting on the state $\ket{+M}=\frac{1}{\sqrt{|x|}}\sum_m\ket{m}^{B_\ems}$.

\begin{align}
\ket{\Psi_{4}}=\sum_{i,k}\alpha_{i,k}\ket{i,k}^{A,B}\ket{q=gx_{i-\mu}y_k}^\emw=e^{ig\Omega_\mu P^\emw}\ket\psi^{A,B}\ket{q=0}^\emw,
\end{align}

where we traced out $\ems$.

\subsection{Remarks}
\begin{itemize}
\item We chose $\ems$ with a discrete Hilbert space, but did not make any restrictions on $\emw$ or on the initial state $\ket{q=0}$.

\item The strong limit appears when $\ket{q=gx_{i-\mu}y_k}$ are almost orthogonal to each other for different values of $gx_{i-\mu}y_k$. That is, $\braket{q=gx_{i-\mu}y_k}{q=gx_{j-\mu}y_l}\approx\delta(x_{i-\mu}y_k-x_{j-\mu}y_l)$, for all $x_{i-\mu},x_{j-\mu}$ in the spectrum of $X$ and $y_k,y_l$ in the spectrum of $Y$.

\item Likewise the weak limit appears when the states  $\ket{q=gx_{i-\mu}y_k}$ completely overlap. That is, $\braket{q=gx_{i-\mu}y_k}{q=gx_{j-\mu}y_l}\approx1-f(x_{i-\mu}y_k-x_{j-\mu}y_l)$ for all $x_{i-\mu},x_{j-\mu}$, with $f(0)=0$ and $f(x_{i-\mu}y_k-x_{j-\mu}y_l)<<1$ for all $x_{i-\mu},x_{j-\mu}$ in the spectrum of $X$ and $y_k,y_l$ in the spectrum of $Y$ (see \cite{Salvail2013} for a more complete derivation of the requirements on the meter states).

\item The permutation $x_{k}\rightarrow x_{k-\mu}$ does not change the spectrum of eigenvalues, i.e. the spectrum of $\Omega_\mu$ is the same as $XY$. However the degenerate subspaces usually change as a result. For example, when $XY$ is the operator $\Pi_{1,1}$, a wrong result on Alice's side will induce a measurement of the operator $\Pi_{0,1}$ instead. This operator has the same spectrum $\{1,0,0,0\}$ but a different eigenstate for the non-degenerate eigenvalue $1$.

\item  In some cases the permutation $x_{k}\rightarrow x_{k-\mu}$ does not make a major difference in terms of the degenerate eigenspaces, i.e. the operators $XY$ and $\Omega_\mu$ have the same degenerate subspaces, with different corresponding  eigenvalues.  For instance, if $X$ is Pauli operator, its eigenvalues are $\pm1$ so a `wrong' outcome on Alice's side will only give an overall $-$ sign in the shift, but the degenerate subspaces will remain the same. In these cases the measurement succeeds regardless of Alice's measurement outcome, although Bob cannot interpret his measurement result without Alice's help (see example below).

\end{itemize}

\subsection{Example 2. A product of Pauli operators}\label{example2}

To induce the Hamiltonian  $\sigma_z^A\sigma_z^BP^{\emw}$ we can use the procedure outlined in the main text (see Fig. 2) and replace Bob's C-C-W with $e^{ig\sigma_z\sigma_zP^\emw}$. It is, however, instructive to follow the procedure above carefully.

We use $X=\sigma_z^A$, $Y=\sigma_z^B$, so  $K_0=\ket{1}\bra{1}$,  $K_1=\ket{0}\bra{0}$ and $\Omega_0=\sigma_z^A\sigma_z^B$, $\Omega_1=-\sigma_z^A\sigma_z^B$. The state $\ket{+_x}^{A_\ems,B_\ems}=\frac{1}{\sqrt{2}}[\ket{00}+\ket{11}]$ and $Q^{B_\ems}\ket{0}=\ket{0}$,  $Q^{B_\ems}\ket{1}=-\ket{1}$.

Alice's steps are simple enough and the state $\ket{\Psi_2}$ is the same as Eq. \eqref{eq:p2} so we continue with Bob's steps
\begin{align}
\ket{\Psi_{3}}=\sum_{i,k}\alpha_{i,k}\ket{i,k}^{A,B}\ket{\mu+i}^{B_\ems}\ket{q=g(-1)^{(\mu+i+k) }}.
\end{align}

Here we used the fact that $\ket{\mu-i}^{B_\ems}=\ket{\mu+i}^{B_\ems}$ (since the Hilbert space is two dimensional) and the relation $\sigma_z^A\sigma_z^B\ket{i,k}^{A,B}=(-1)^{i+k}\ket{i,k}$.

The erasure step is based on a measurement of $\sigma_x^{B_\ems}$.  With the result $+1$ we will get

\begin{align}
\ket{\Psi_{4}}&=\sum_{i,k}\alpha_{i,k}\ket{i,k}^{A,B}\ket{q=g(-1)^{(\mu+i+k) }}\\
&=e^{(-1)^\mu ig\sigma_z^A\sigma_z^B P^\emw}\ket\psi^{A,B}\ket{q=0}^\emw.
\end{align}

While a result $-1$ will yield

\begin{align}\label{psi4tilde}
\ket{\widetilde{\Psi_{4}}}&=\sum_{i,k}(-1)^{2\mu+i}\alpha_{i,k}\ket{i,k}^{A,B}\ket{q=g(-1)^{(\mu+i+k)}}\\
&= \sigma_z^{A}e^{(-1)^\mu ig\sigma_z^A\sigma_z^B P^\emw}\ket\psi^{A,B}\ket{q=0}^\emw.
\end{align}

Overall, Bob will register the correct result up to a possible $-$ sign, which he can correct later by exchanging information with Alice. However, the state could have changed on Alice's side. This can also be corrected, but only when Alice receives information from Bob. This correction does not require communication  in the strong case (see below).

\subsubsection{The strong limit}

  In the case of a strong measurement, Bob can correct $\ket{\widetilde{\Psi_{4}}}$ locally since he effectively has access to it. By applying $e^{i\frac{\pi}{2}\sigma_z^BQ^\emw}$ to $\ket{\widetilde{\Psi_{4}}}$  in Eq. \eqref{psi4tilde} he  arrives at

\begin{align}
\ket{\widetilde\Psi_5}&=\sum_{i,k}(-1)^{i}\alpha_{i,k}e^{i\frac{\pi}{2}\sigma_z^B(-1)^{(\mu+i+k)}}\ket{i,k}^{A,B}\ket{q=g(-1)^{(\mu+i+k)}}\\
&=\sum_{i,k}(-1)^{i}\alpha_{i,k}e^{i\frac{\pi}{2}(-1)^{(\mu+i+2k)}}\ket{i,k}^{A,B}\ket{q=g(-1)^{(\mu+i+k)}}\\
&= i(-1)^\mu\sum_{i,k}\alpha_{i,k}\ket{i,k}^{A,B}\ket{q=g(-1)^{(\mu+i+k)}}=ie^{g\sigma_z^A\sigma_z^BP^\emw}\ket{\Psi_4}\ket{q=0}.
\end{align}

Hence, the  procedure is deterministic.

\subsubsection{The weak limit}
While the procedure can be deterministic at the strong limit, it is impossible to make it deterministic in general. The  induced interaction should be  $U=e^{i g \sigma^A\sigma^B P^\emw}$. Now let us assume this could be done for any initial state of the meter in a deterministic way. In such a case Bob could prepare the meter in the state $\ket{p=1}^\emw$ with $P^\emw\ket{p=1}^\emw=\ket{p=1}^\emw$, so $U\ket\psi^{A,B}\ket{p=1}= e^{i g \sigma^A\sigma^B}\ket\psi^{A,B}\ket{p=1}^\ems$, but if this could be done instantaneously then it would have been possible to send superluminal signals between Alice and Bob.

\subsubsection{Relation to other works}
The products of Pauli operators on specelike seperated systems appear in a large number of works. One example is the CHSH operator \cite{PhysRevLett.23.880} which is a sum of four product Pauli terms. It is easy to modify our scheme to measure a sum by using standard methods for measuring sums of operators \cite{Brodutch2009}. We note, however, that  such a measurement would not be used to violate the CHSH inequality in the usual way. A more direct approach for using our method  would be to improve the results of Higgins et al.  \cite{Higgins2015}. They used local weak measurements with post-selection to measure the CHSH operator directly in an experiment. However, at each run they could only measure three out of the four terms. By measuring non-local weak values it would be possible to measure all four terms simultaneously.

A second use of product terms is in stabilizer measurements  \cite{Beckmanetal2001,Gottesman2007}. In this case the subsystems are technically close enough to each other and the measurements are local. However in practice it is often difficult to generate (or engineer) the interaction terms. Methods used to tackle this problem (e.g \cite{Ghosh2012}) are often based on non-local measurement techniques such as modular measurements. It is an open question  whether  the  erasure method would provide any advantage in these situation.

\section{Comparison with other methods}\label{AppD}

\subsection{Weak measurements beyond the von Neumann formalism}

Outside the von Neumann model  it  is  still unknown which weak values are directly observable, and of those, which are observable via weak measurements \cite{BrodutchMsc}. The erasure scheme allows a very wide range of observables to be measured in a weak measurement by creating an effective von Neumann Hamiltonian. To compare the erasure scheme with other weak measurement schemes, that deviate from the usual von Neumann model, we will define the weak measurement as the weak limit of the general measurement protocol described below.

 The measurement protocol has a variable strength parameter $g$ and involves  interaction between the measured system $\es$ and a meter $\emw$. The meter is characterized  by a pointer observable  $Q$ and conjugate momentum $P$.  The outcome of the measurement is the change in the meter's state at the end of the protocol. Generally the initial state of the meter, which we label   $\ket{0}^\emw$,  can  depend on the interaction strength $g$ which varies from $g=0$ (no measurement) to $g=1$ (strong measurement).

The measurement can be described by a superoperator $\eW$ taking a pure product system-meter state $\rho^{\es\emw}_0=\ket{\psi}\bra{\psi}\otimes\ket{0}\bra{0}$ to a final state $\eW(\rho^{\es\emw}_0)$. The superoperator  $\eW$ changes smoothly as we vary  $g$ and the  following properties hold at the weak limit  $g<<1$:
{\prop {\bf - Non-disturbing} - The probability of post-selecting a state $\ket{\phi}$, given by $P(\phi)=tr[\bra{\phi}\eW(\rho^{\es\emw}_0)\ket{\phi}]$,  is unaffected by the measurement up to terms of order $g^2$
\begin{equation}
P(\phi)=|\braket\phi\psi|^2[1-O(g^2)]
\end{equation}
}
\vspace{-10pt}
{\prop {\bf - Weak potential} - After measurement and post-selection, the meter's state is shifted by a value proportional to the weak value
\begin{equation}
\{\Os\}_w=\frac{\bra\phi \Os \ket\psi}{\braket\phi\psi}.
\end{equation}
so that
\begin{align}
\bra{\phi}\eW(\rho^{\es\emw}_0)\ket{\phi}= e^{-igH_{w}}\ket{0}\bra{0} e^{igH_{w}}+O(g^2).
\end{align}}
and  $H_w\propto \W{\Os}P$.\\

We also require that when $g=1$ we have a strong measurement
{{\prop {\bf - Strong limit} - $\mathcal{W}_1(\rho^{\es\emw}_0)$ is a strong (von Neumann) measurement.\\}

 Property 1 allows us to make claims about the measurement without resorting to counterfactual statements, e.g. by making simultaneous measurements of incompatible observables.   Properties 2 and 3 are necessary to interpret the weak value as the result of a measurement \cite{CommentAB,CommentEC}. An operational variant of these three properties allowed Pusey \cite{Pusey2014} to build a model showing \emph{anomalous weak values are proofs of contextuality}.

\subsection{Comparison with other non-local weak measurement methods}

Hardy's paradox  provided some motivation for the previous attempts at finding a scheme for non-local weak measurements.  Resch and Steinberg (RS) showed that it is possible to \emph{extract} $\{\Pi_{m,n}\}_w$ from  local weak measurements \cite{Resch2004a}. Their method was later demonstrated experimentally and inspired other intriguing works \cite{Mitchison2007,Lundeen2005,Lundeen2009}.  Kedem and Vaidman showed that modular values \cite{Kedem2010} provide a more efficient method to extract  weak values of a subset of unitary observables. In the case of Hardy's experiment the non-local observables are not of this form but they can be  calculated from local and non-local modular values as was done in a preceding  experiment \cite{Yokota2009}.

Both methods, although inspiring,  fail the required  properties above \cite{Brodutch2009,Kedem2010}, and therefore cannot be completely considered as weak measurements. Both measurements are not a weak version of a strong measurement and therefore immediately fail property 3. In the RS scheme there is no single meter that contains the measurement results and  the relevant correlations are only apparent at second order (in $g$), so properties 1 and 2 are also missing. Moreover, from an experimental perspective this scheme is difficult to implement since the  resources required to observe the correlations are of order $g^2$ \cite{Brodutch2009}.

The KV modular value protocol \cite{Kedem2010} can be tuned in such a way that it follows a modified version of property 2 for a limited class of observables.  The protocol   can be summarized as follows: The meter is a qubit initially in the state $\alpha\ket{0}+\beta\ket{1}$
and the interaction Hamiltonian is $H_I\propto\ket{1}\bra{1}\otimes\tilde O$, where $\tilde O$ is an Hermitian operator on the system to be measured. 
 After post-selection the state of the meter is $\alpha\ket{0}+\beta\{O\}_M\ket{1}$, where $\{O\}_M=\frac{\bra{\phi}e^{iK\tilde O}\ket{\psi}}{\braket{\phi}{\psi}}$ is the modular value and $K$ is a real parameter.

The modular value is independent of the strength of the measurement, which is proportional to $\beta$. For  $\beta\{O\}_M<<1$ we get the effective Hamiltonian $H=-iK\ket{1}\bra{1}\log\{O\}_M$, giving a  slightly modified version of property 2, {\it i.e.}, the meter's shift is proportional to $K \log\{O\}_M$ rather than  $ \{O\}_M$.

Unlike weak values,  modular  values   obey a product rather than a sum rule, with respect to $\tilde O$. For example, the modular value of a sum of single qubit Pauli operators  is the same as the weak value of the product of these operators\cite{Kedem2010}. 
An operational advantage is that the modular values do not require a weak value approximation, they can be measured for any value of $\beta$,  making them easier to measure in an actual experiment. On the other hand, the modular values are limited to unitary operators of the type $e^{i\tilde O}$.

\subsection{Comparison with other strong non-local measurement methods}

Before comparing with other methods having only a strong limit, we note that the only known non-local observables that can be measured in a scheme that has both strong and weak limits are sum observables of the type $A+B$ \cite{Brodutch2009}. The technique used is similar to the more general method of measuring a modular sum $(A+B)mod~k$, where $k$ is some fixed integer. These modular  measurements can also be used for measuring some types of product observables, e.g. the modular sum $(\sigma_z^A+\sigma_z^B) mod~2 = \sigma_z^A\otimes\sigma_z^B$. Nevertheless, the class of observables measured in this way remains limited.

 The modular sum however, has no weak limit. Seemingly, the issue with a weak limit is related to the fact that the meter can only be modular for a fixed strength measurement \cite{BrodutchMsc}. The problem is, however, deeper and related to causality. In general, if the observable $\Omega^{AB}$ generates some interaction between Alice and Bob,  the back action of a weak measurement can be used for signalling. At the strong limit the ability to send signals disappears due to the modular structure of the observable. In example 2 above we can see that this carries over to the erasure scheme, i.e. the observable $(\sigma_z^A+\sigma_z^B) mod~2 = \sigma_z^A\otimes\sigma_z^B$ can be measured deterministically.

The more general method for measuring non-local observables is Vaidman's scheme \cite{Vaidman2003}, which involves an infinite number of partial teleportation rounds. That scheme is not only extremely costly in terms of resources \cite{Clark2010}, it is also destructive. Although the right outcome is obtained at the readout stage, the measurement distributes the information encoded in the initial system to all the entangled pairs, such that the system's state at the output is independent of the measurement result and/or the input state. Moreover, the scheme is adaptive although it does not require communication.

It is simple to turn Vaidman's  partial  teleprtation scheme into a non-deterministic L\"{u}ders measurement scheme. One can simply stop after the second round of half teleportations and post-select on the desired results. This is always more expensive than the erasure scheme in terms of success probability and entanglement resources.  It requires two successful teleportation rounds, whereas the worst case erasure schemes require the same resources as a single teleportation round.

\section{The erasure scheme beyond non-local measurements}\label{AppE}

The erasure scheme can be used in practical applications that go beyond the realm of non-local measurements.  In the main text we used this scheme to construct a three-body  von Neumann Hamiltonian with a restriction on the possible allowed interactions. While this Hamiltonian is fundamental in  measurement theory, it is in fact very general and has a variety of potential uses, in particular with respect to quantum information processing. Moreover, the erasure scheme can be used to construct more general, many-body interaction Hamiltonians, under a variety of constraints on the allowed interactions.   In this section we present one example application of the erasure scheme for constructing  a generic Controlled-Controlled-Unitary interaction  with  tunable parameters $P^\emw$ and  $g=\int_0^\tau f(t) dt$ that define the target unitary. 

Our   example is based on  the three body Hamiltonian related to the measurement of $\ket{11}\bra{11}$ discussed previously. The  restriction on non-local interactions is replaced by  a different  realistic restriction.   In particular, we consider  a scenario  where it is only possible to engineer specific one and two body  terms in the Hamiltonian. These restrictions have a similar structure to the restrictions imposed by relativistic causality, although they may arise for different reasons ranging from fundamental constraints (e.g. photons do not  interact directly) to specific constraints related to  particular architectures (e.g. constraints due to geometry).  Consequently the erasure scheme is a useful tool in these scenarios.  Our goal is to construct  the Hamiltonian that generates a Controlled-Controlled-Unitary gate under the restrictions that the subsystems involved cannot interact directly and must do so using an ancilla.  One advantage of this   scenario is that the erasure step can be deterministic and may not even require a correction step.

To keep the discussion simple and general, we use abstract terminology. However, it is possible to think of the subsystems of $\es$ as qubits encoded on the path of two photons such that the computational basis states $\ket{0}^A,\ket{1}^A$ correspond to left and right paths $\ket{L}^1,\ket{R}^1$ of photon 1 respectively and likewise for $B$ with photon 2. Similarly, $\emw$  can be a qubit encoded in the path of a third photon with the computational basis states $\ket{0}^\emw,\ket{1}^\emw$ encoded in superposition states $\frac{1}{\sqrt{2}}\left(\ket{L}^3+\ket{R}^3\right), \frac{1}{\sqrt{2}}\left(\ket{L}^3-\ket{R}^3\right)$. The  meter, $\ems$, can be an atomic system that can interact with the photons in a nondestructive way. Such systems were demonstrated in \cite{Reiserer2013}. Our assumption is that the photons cannot interact with each other and all interactions must be mediated via $\ems$. 

\subsection{Example: A controlled-controlled-unitary}

Consider the  measurement of the  observable $\Omega^\es=\Pi_{1,1}^\es=\ket{11}\bra{11}^\es$ under the restriction that qubits can only interact with an ancila $\ems$ (i.e. they cannot interact directly). We want the result of the measurement to register on a third qubit $\emw$.   The interaction Hamiltonian associated with this observable, 
\begin{equation}\label{intham}
H_I=f(t) \ket{1}\bra{1}^A\otimes \ket{1}\bra{1}^B\otimes P^\emw,
 \end{equation}
 cannot be implemented directly. Note that although we use the term \emph{measurement},  the Hamiltonian above represents a Controlled-Controlled-Unitary gate with $U=e^{igP^\emw}$. For $g=\int_0^\tau f(t) dt=\frac{\pi}{2}$ this is similar to a  Toffoli gate while for more general choices it corresponds to a Deutsch gate which is universal for quantum computing \cite{NielsenandChuang,Deutsch1989}.  If one allows only two-body interactions it is well known that the controlled-controlled-U gate can be constructed using five two body interaction gates \cite{NielsenandChuang}. Three of these gates depend on the specifics of $U$, in particular for  $U=e^{igP^\emw}$, these three gates will depend on  the choice of both $g$ and $P^\emw$. Our requirements are, however, more stringent: First, we would like $g$ and $P^\emw$   to  appear as few times as possible in the implementation  (the scheme below requires only one gate that depends on these); second, we do not allow arbitrary two body interactions, and only allow interactions that are mediated by the ancilla. A photonic Toffoli gate following  similar restrictions on the interactions was  recently  designed in \cite{Luo2015}, using a technique which is based on the erasure method, i.e.  each two qubit gate is mediated via an ancilla that gets erased. Our scheme below is similar to that scheme but has a few distinct advantages: It can be performed deterministically without  correction steps (that are usually hard  to realize),  it requires only a single erasure step (as opposed to five),  it does not require an ancillary photon to mediate the interactions and it allows us to freely choose  $g$.

\subsubsection{The scheme}

Prepare $\ems$, a qutrit,  in the state $\ket{0}^\ems$ and  take $P^\ems\ket{i}=\ket{i+1}$. The meter is prepared in the state $\ket{0}^\emw$  with $P^\emw=\sigma_x^\emw$. 

The initial state is 
\begin{equation}
\ket{\psi}^\es\ket{0}^\ems\ket{0}^\emw=\sum_{k=0,l=0}^{1,1}\alpha_{k,l}\ket{kl}^\es\ket{0}^\ems\ket{0}^\emw
\end{equation}

We now let each subsystem $A,B$ interact with $\ems$ via a controlled-$P^\ems$, and then let $\emw$ interact with $\ems$ via 
\begin{equation}\label{finalint}
\left[\ket{0}\bra{0}^\ems+\ket{1}\bra{1}^\ems\right]\otimes\openone^\emw+\ket{2}\bra{2}^\ems\otimes e^{igP^\emw}
\end{equation}

The state will then be 
\begin{equation}
\left[\alpha_{0,0}\ket{00}^\es \ket{0}^\ems+(\alpha_{0,1}\ket{01}+\alpha_{1,0}\ket{10})\ket{1}^\ems\right]\ket{0}^\emw+\alpha_{1,1}\ket{11}^\es\ket{2}^\ems e^{igP^\emw}\ket{0}^\emw
\end{equation}

We can now erase $\ems$ by using the  erasure protocol described in the main text and succeed with probability $1/3$. However, since the setup is local we can use a deterministic  method to erase the information. We  couple each system qubit  to $\emw$ via a Controlled-$[P^\emw]^\dagger$ and reverse the first two interactions.  In either case we construct the effective three-body Hamiltonian of Eq. \eqref{intham}.  Importantly, $P^\emw$ and  $g$ appear  only in one gate, Eq. \eqref{finalint}.

\twocolumngrid

\end{document}